\documentclass[aps,preprint,groupedaddress,floatfix]{revtex4}

\usepackage{epsfig}
\newcommand{\p}{\partial_PH_S}
\newcommand{\wm}{\omega_{0}}
\begin{document}

\title{Semiclassical theory of energy diffusive escape in a Duffing oscillator}

\author{Alvise Verso and Joachim Ankerhold}

\affiliation{Institut f\"ur Theoretische Physik, Universit\"at Ulm, Albert-Einstein-Allee 11,
89069 Ulm, Germany}

\date{\today}

\begin{abstract}
Motivated by recent experimental progress to read out quantum bits implemented in superconducting circuits via the phenomenon of dynamical bifurcation, transitions between steady orbits in a driven anharmonic oscillator, the Duffing oscillator, are analyzed. In the regime of weak dissipation a consistent master equation in the semiclassical limit is derived to capture the intimate relation between finite tunneling and reflection and bath induced quantum fluctuations. From the corresponding steady state distributions analytical expressions for the switching probabilities are obtained. It is shown that a reduction of the transition rate due to finite reflection at the phase-space barrier is overcompensated by an increase due to environmental quantum fluctuations that are specific for diffusion processes over dynamical barriers. Moreover, it is revealed that close to the bifurcation threshold the escape dynamics enters an overdamped domain such that the quantum mechanical energy scale associated with friction even exceeds the thermal energy scale.
\end{abstract}

\maketitle

\section{Introduction}
The prospect to tailor devices for quantum information processing  has stimulated major experimental research in the last years. Several different technologies have been explored to assess the possibility to realize quantum bits, like ion traps \cite{monz_10}, liquid state magnetic resonance, linear optics, electrons in liquid helium and superconducting Josephson Junction (JJ) devices \cite{vion_02,claudon_04,koch_07}. Particularly for the latter ones the insulation of the structure in which the quantum bit is implemented, e.g.\ a Cooper pair box, from its surrounding is of substantial relevance. This issue also includes the readout device of the qubit state which must be designed such as to minimize its presence on the one hand, but to efficiently gather the required information on the other hand.

 A powerful readout scheme is based on the phenomenon of dynamical bifurcation, realized in form of the Josephson Bifurcation Amplifier \cite{siddiqi_04,siddiqi_05} and the Cavity Bifurcation Amplifier \cite{vion_02}. A big JJ is placed in parallel to a Copper pair box and driven my an external microwave source. Accordingly, close to the first bifurcation threshold determined by the frequency and the amplitude of the drive, two stable oscillations appear in the big JJ with thermal fluctuations inducing transitions between them. The sensitivity of this process to the shape of the Josephson potential
 is used to retrieve information about the qubit state.
 However, the possibility  to tune parameters over wide ranges make these systems interesting on their own as devices to study fundamental aspects of driving, nonlinearity and dissipation. In particular, the question about the impact of quantum fluctuations on transitions between two stable basins of attraction in phase space goes far beyond the standard situation for escape over static energy barriers.

Theoretically, within the relevant range the driven big JJ can be described as a Duffing oscillator \cite{manucharyan_07}. This oscillator is particularly important because it represents the simplest model to analyze phenomena like bifurcation, period doubling, and dynamical tunneling.
The classical dynamics of this system is well known and the transition between the two stable states, induced by thermal fluctuations,  has been investigated in detail \cite{dykman_79,dykman_80}.
With lowering temperature quantum mechanical effects appear. It is known that their  contributions are twofold. On the one hand, the transmission probability through the barrier becomes finite, on the other hand quantum fluctuations of the environment appear due to finite zero point fluctuations in the well adjacent to the barrier and to quantum fluctuations in the diffusion coefficients. For escape over static barriers it turns out that for weak dissipation the former effect may even exceed the latter one to produce a reduction of the escape rate compared to the classical situation \cite{verso_09} due to a finite reflection of states above the barrier energy. 
 Thus, the question arises whether the same is true for transitions over dynamical barriers.
Moreover, it has to be explored how a consistent semiclassical description for driven dissipative anharmonic systems must be formulated. To solve both issues is the purpose of this paper.

A  powerful procedure for analytical investigations is to describe driven oscillators in a frame rotating with a frequency equal to the response frequency of the system \cite{dykman_80,dykman_98}. Using this approach Dykman and co-workers analyzed the diffusive escape for the Duffing oscillator \cite{dykman_88} and for other periodically driven systems \cite{marthaler_06}. The focus there has been on reservoir induced quantum effects. Complementary,
 in \cite{serban_07}  macroscopic  quantum tunneling has been addressed in the deep quantum regime. A numerical description of the problem, taking into account multiphoton resonances, has been given in \cite{peano_06}. In the present work we use a semiclassical approach in order to calculate systematically quantum corrections to the classical escape rate.
We consider both mechanisms, tunneling and bath fluctuations, starting from a properly derived master equation. In contrast to \cite{dykman_88,marthaler_06,serban_07} the impact of finite barrier reflection/tunneling and the whole structure of the dissipative dynamics in the rotating frame \cite{verso_10} are included in this master equation.
The corresponding analytical expressions for the escape rate apply to the range of weak damping and moderate temperatures.

The paper is organized as follows: In Sec.~\ref{sec2} we introduce the model and the basic notation including the mapping to the rotating frame. This description is extended in Sec.~\ref{sec3} to explicitly include also the bath degrees of freedom. This formulation provides the basis to derive in Sec.~\ref{sec4} a semiclassical expansion of the master equation the steady state distributions of which are used to derive the escape rates in Sec.~\ref{sec5}. In Sec.~\ref{sec4} this quantum diffusion equation is discussed and the quantum corrections to the classical escape rate are obtained.

\section{System and mapping on a rotating frame}\label{sec2}
We consider a system with a weakly anharmonic potential driven by an external time-periodic force (Duffing oscillator), namely
\begin{equation}\label{6:h}
H_S(t)=\frac{1}{2 M} p^2+\frac{1}{2}M \wm^2q^2-\frac{1}{4}\Gamma q^4+Fq \cos(\omega_{ d} t)\,.
\end{equation}
Accordingly, for the anharmonic coefficient, we assume $\Gamma\langle q^2\rangle\ll M\omega_0^2 $ so that driving is almost resonant for
\begin{eqnarray}
\delta\omega=\omega_0-\omega_{ d}\ll \omega_{ d}\,.
\end{eqnarray}
Classically, when damping is taken into account, two stable oscillations with different amplitudes and phases appear beyond a bifurcation threshold. The latter one depends on external parameters such as driving amplitude $F$ and frequency mismatch $\delta\omega$. In phase space, these two stable states correspond to stable basins of attraction which are separated by an unstable domain. Thermal fluctuations may induce transitions between these basins that in turn carries information about the global shape of the phase space barrier and the environment.

\begin{center}
\begin{figure}
\centering
\epsfig{file=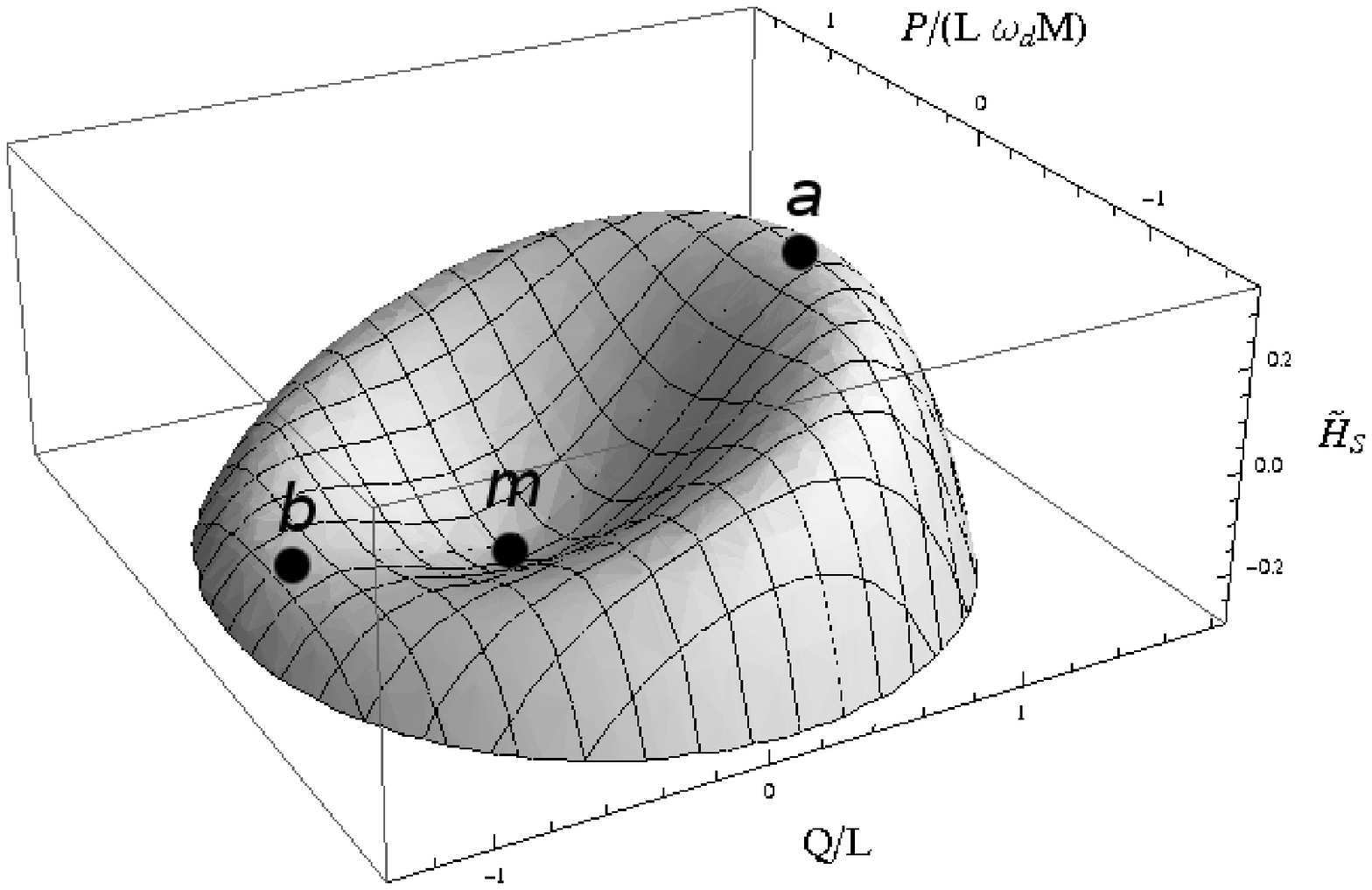, height=6cm}
\epsfig{file=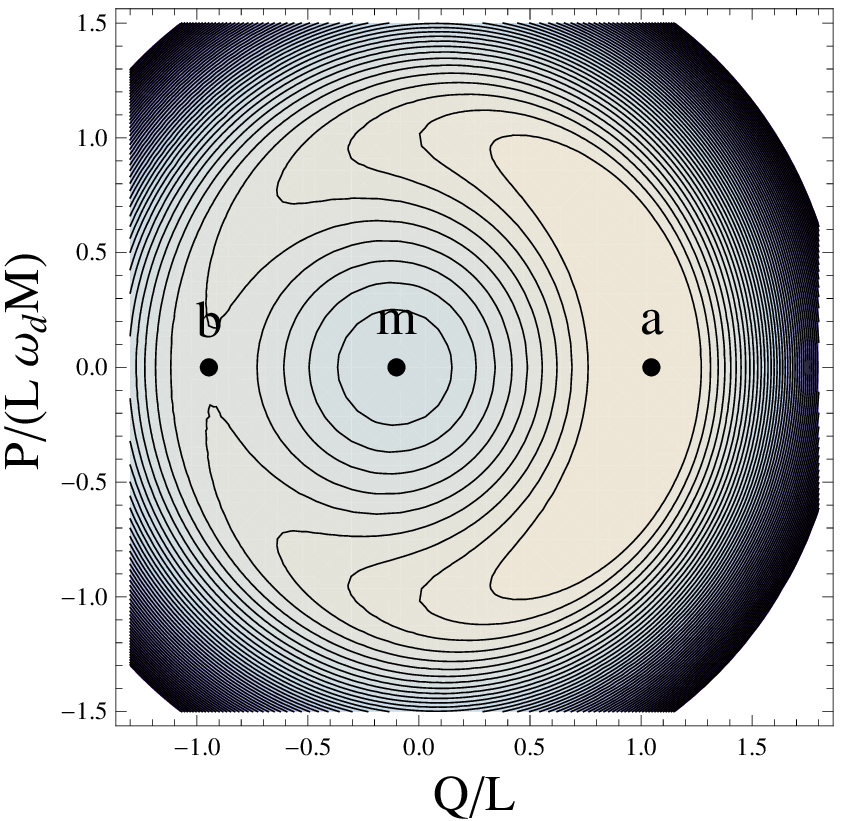, height=6.5cm}
\caption{The Hamiltonian function (from Eq.(\ref{6:hs})) in the rotating frame, for $\alpha=1/27$. The energy is scaled with $ \omega_{ d}M\delta\omega L^2$. The minimum (m) and the maximum (a) in the figure are the stable sates with low and high amplitude, respectively, separated by a marginal state (b).}\label{6:fig_pot}
\end{figure}
\end{center}

\begin{center}
\begin{figure}
\centering
\epsfig{file=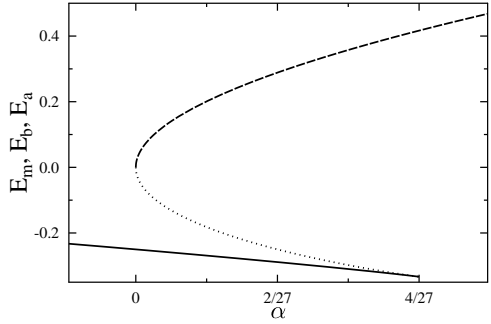, height=6cm}
\caption{The energies $E_m$ (solid), $E_b$ (dotted) and $E_a$ (dashed)  of the minimum, the marginal state and of the maximum as functions of $\alpha$. The barrier hight $\Delta V\equiv E_b-E_m$ is maximal for $\alpha=0$ and vanishes for $\alpha=4/27$. The energies are scaled with $L^2\omega_{ d}M\delta\omega$.}\label{6:fig_en}
\end{figure}
\end{center}

Theoretically, the difficulty for a rate description in this kind of system is that the  Hamiltonian of the isolated system $H_S(t)$ is time-dependent and, therefore, energy is not conserved.
However, the dissipative system approaches a steady-state situation such that the reduced density matrix takes the form $\rho(t)\sim \bar{\rho}(t)\cos(\omega_{ d} t)$ with an only weakly time-dependent density $\bar{\rho}$. For the further analysis it is thus convenient to switch to a rotating frame, given by the unitary operator,
\begin{equation}
 U_S(t)=e^{-i\hat{a}^{\dag}\hat{a}\omega_{ d}t},
\end{equation}
where $\hat{a}=\sqrt{\frac{2M}{\hbar\omega_{ d}}}\left(\omega_{ d}q+\frac{i}{M}p\right)$ and $\hat{a}^{\dag}=\sqrt{\frac{2M}{\hbar\omega_{ d}}}\left(\omega_{ d}q-\frac{i}{M}p\right)$ are the annihilation and creation operators for {\em harmonic} oscillators in the system, respectively.
The transformation $ U_S(t)$ applied to the coordinate $q$ and momentum $p$ give
\begin{eqnarray}
U_S^{\dag}\,q\,U_S&=&Q\cos(\omega_{ d}t)+\frac{1}{\omega_{ d} M}P\sin(\omega_{ d}t),\\
U_S^{\dag}\,p\,U_S&=&-\omega_{ d} MQ\sin(\omega_{ d}t)+P\cos(\omega_{ d}t)\,
\end{eqnarray}
with $Q$ and $P$ as new (slowly varying) coordinates.
From these equations it is clear that the unitary transformation is the equivalent to a rotation of the classical phase space.
In the rotating frame determined by $U_S(t)$ the Hamiltonian reads
\begin{eqnarray}\label{6:hs}
\tilde{H}_S&=&U_S^{\dag}\left[H-i\hbar\frac{\partial}{\partial t}\right] U_S
\\
&=&M\omega_{ d}\,\delta\omega\, L^2\left[-\frac{1}{4}\left(\frac{Q^2}{L^2}+\frac{P^2}{(L\omega_{ d} M)^2}-1\right)^2+\frac{\sqrt{\alpha}}{L}Q\right]\nonumber
\end{eqnarray}
with a length scale
$L=\sqrt{\frac{8\omega_{ d} \delta\omega M}{3\Gamma}}$ and the  bifurcation parameter
\begin{equation}\label{bifur}
\alpha=\frac{3F^2\Gamma}{32(\omega_0\delta\omega M)^3}\, .
\end{equation}
 Moreover following a rotating wave approximation fast oscillating terms $\exp(\pm in\omega_{ d} t)\,$, $n\geq1$ are neglected such that a time independent Hamiltonian $\tilde{H}_S$ is obtained.
 For $0<\alpha<4/27$ the rotating frame system exhibits three extrema, whereas the two stable ones correspond in the laboratory frame to oscillations with low and high amplitude, respectively. They are separated by a phase-space barrier associated with an unstable extremum (see fig.\ref{6:fig_pot} and fig.\ref{6:fig_en}).
For $\alpha<0$ only the low amplitude states exist and for $\alpha>4/27$  only the high amplitude state remains.

\section{System and bath: microscopic description}\label{sec3}
In order to describe quantum dissipation, we explicitly introduce a bath coupled to the system, so that the total Hamiltonian is given by
\begin{eqnarray}\label{6:ht}
H  = H_S + H_R + H_I
\end{eqnarray}
 with a system part as in (\ref{6:h}), a reservoir part $H_R$ and an interaction $H_I$, i.e. \cite{weiss_07,caldeira_83}
\begin{eqnarray}
\label{eq1}
 H_R+H_I & =& \sum_\alpha \frac{p^2_\alpha}{2m_\alpha} + \frac{m_\alpha \omega^2_\alpha}{2} \left(x_\alpha- \frac{c_i}{m_\alpha\omega_\alpha^2}  q \right)^2 \, .
\end{eqnarray}
We switch to the rotating frame with the unitary operator of the composite system \cite{serban_07,verso_10},
\begin{equation}
\label{eq:unitary}
U(t)\equiv U_S(t)U_R(t)=e^{-i\hat{a}^{\dag}\hat{a}\omega_{ d} t-i \sum_n^N\hat{b}_n^{\dag}\hat{b}_n\omega_{ d} t}\,,
\end{equation}
whereas $\hat{b}_n$ and $\hat{b}_n^{\dag}$ are annihilation and creation operators for harmonic oscillators in the bath. In the rotating frame the total Hamiltonian reads
\begin{eqnarray}\label{eq:h}
\tilde{H}=\tilde{H}_S+\tilde{H}_R+\tilde{H}_{I}\,
\end{eqnarray}
with $\tilde{H}_S$ as in (\ref{6:hs}) and
\begin{eqnarray}\label{6:hr}
\tilde{H}_R&=&\sum_{n=1}^N\frac{{p}_n^2}{2\tilde{m}_n}+\frac{\tilde{m}_n}{2}\tilde{\omega}_n^2 {x}_n^2,\nonumber\\
\tilde{H}_{I}&=&-\sum_{n=1}^N \tilde{c}_n\left({x}_n {Q}+\frac{{p}_n}{\tilde{\omega}_n\tilde{m}_n}\frac{P}{\omega_{ d}M}\right)\nonumber\\
&+&\left({Q}^2+\frac{P^2}{(\omega_{ d} M)^2}\right)\sum_{n=1}^N\frac{{c}^2_n}{4 m_n \omega_n^2}\, ,
\label{introt}
\end{eqnarray}
where the new bath parameters read
\begin{eqnarray}
\tilde{m}_n=\frac{m_n}{1-\omega_{ d}/\omega_n}\, ,\ \tilde{\omega}_n=\omega_n-\omega_{ d}\, , \ \tilde{c}_n=\frac{c_n}{2}\,.
\end{eqnarray}

With the unitary transformation (\ref{eq:unitary}) the total Hamiltonian $H$ in the laboratory frame (\ref{6:ht}) is mapped onto a new  Hamiltonian $\tilde{H}$ in the moving frame, composed of the system part (\ref{6:hs}), a reservoir part (\ref{6:hr}) with new  parameters and an  interaction part (\ref{introt}). The mapped composite system can now be described by techniques applied for undriven escape problems with the notable difference though, that the interaction  between system and environment becomes more complex containing in addition the the conventional position-position coupling also momentum-momentum contributions. However, as shown in \cite{verso_10}, following the standard procedure all bath properties can be captured by a spectral density and by temperature.
Consequently, in the rotating frame we can adapt the approach developed in \cite{verso_09} for  transition rates over energy barriers in the energy diffusive limit (weak dissipation).

One starts from the time evolution of the density matrix of the full compound $W(t)$ obeying the Liouville-von Neumann equation $i\hbar d{W}(t)/dt=[\tilde{H},W(t)]$ with an initial state $W(0)$.
The relevant operator is the reduced density $\rho(t)={\rm tr}_B\{W(t)\}$ after eliminating the bath degrees of freedom for which a simple equation of motion does in general not exist. In case of weak friction and sufficiently fast bath modes, however, progress is made within a Born-Markov approximation. One then obtains
a master equation which for the present case can be cast in the form \cite{verso_10}
\begin{equation}
i\hbar\frac{d\rho}{dt}=[\tilde{H}_S,\rho]
+\left({{\cal L}}_{QQ}+\frac{{\cal L}_{QP}}{\omega_{ d}M}+\frac{{\cal L}_{PQ}}{\omega_{ d}M}+\frac{{\cal L}_{PP}}{(\omega_{ d}M)^2}\right)[\rho]\, .
\label{eq:master_rot}
\end{equation}
Here operators ${{\cal L}}_{xy}$  are defined according to
\begin{eqnarray}
{\cal L}_{xy}[\rho]&=&\int_0^{\infty}ds\, K^{\prime}_{xy}(s)\left[x,\left[y(-s),\rho(t)\right]\right]\nonumber\\
&&+\int_0^{\infty}ds\, i K_{xy}^{\prime\prime}(s)\left[x,\lbrace y(-s),\rho(t)\rbrace\right]\,,
\end{eqnarray}
with operators $y(s)$  in the Heisenberg representation and $\{,\}$ denoting the anti-commutator.
In the rotating frame the force-force correlator functions are defined by
\begin{eqnarray}\label{force0}
K_{xy}&=&K'_{xy}+iK''_{xy}\nonumber\\
&=&\frac{1}{\hbar}\langle F_x(t)\,F_y(0)\rangle_\beta\hspace{.8cm}\,\, x,y=Q,P\, ,
\end{eqnarray}
where  the bath forces read according to (\ref{6:hr})
\begin{eqnarray}\label{force}
F_Q=\sum \tilde{c}_n x_n\,,\hspace{1cm}F_P=\sum \tilde{c}_n \frac{p_n}{\tilde{\omega}_n\tilde{m}_n}\,.
\end{eqnarray}

Our goal is now to derive from the above master equation (\ref{eq:master_rot}) a semiclassical equation in the energy diffusive limit to determine the leading quantum corrections (order $\hbar$) to the transition rates between the two phase space basins. In case of no external driving the analysis in \cite{verso_09} revealed that a conventional position-position interaction between bath and system produces in the corresponding diffusion equation only quantum corrections of order $\hbar^2$. The leading impact of quantum mechanics is thus due to finite transmission through the barrier, i.e.\ tunneling and reflection. Here, however, we will see that while the contribution ${\cal L}_{PP}$ has a behavior similar to ${\cal L}_{QQ}$, the unconventional bath contributions ${\cal L}_{QP}$ and ${\cal L}_{PQ}$ yield a supplementary correction of order $\hbar$.

\section{Semiclassical master equation}\label{sec4}
An energy diffusion operator can be derived starting either from a discrete or a continuous spectrum in the well of one the stable domains. The difference is though, that the latter procedure conveniently accounts for barrier tunneling near the barrier top, where the density of states is basically continuous.
It is then convenient to introduce the occupation probability of a well state with energy $E$ via
\begin{equation}
P(E,t)=\sum_{n=0} \delta(E-E_n) p_n(t)\, .
\end{equation}
Here $p_n$ is the occupation probability of a well eigenstate with quasi-energy $E_n$ and identical to the diagonal part of the reduced density matrix in the energy representation. The explicit construction of these states follows a type of WKB-recipe as shown below. The time evolution equation (\ref{eq:master_rot}) can now be represented (see \cite{verso_09}) as
\begin{eqnarray}\label{eq:m}
\dot{P}(E,t)&=&\int_{}dE'\left[W_{E,E'}\frac{R(E^{\prime})P(E',t)}{n(E^{\prime})}\right.\nonumber\\
& &\left.-W_{E',E}
\frac{R(E)P(E,t)}{n(E)}\right]-T(E)\frac{\omega(E)}{2 \pi}P(E,t)\,
\end{eqnarray}
with $\omega(E)$ being the frequency of a classical oscillation at energy $E$ and $n(E)$ the density of states. Equation (\ref{eq:m}) captures the incoming probability flux to and outgoing probability flux from the state $E$ according to
intrawell transition rates \cite{breuer_02}
\begin{eqnarray}\label{eq:Wq}
%W_{g,E'}&=&\frac{1}{\hbar^2}\int_{-\infty}^{\infty}dt {\rm Tr}_R\{\langle E|H_{I}(t)|E'\rangle\langle E'|H_{I}|E\rangle\rho_R^{eq}\}
W_{E,E'}&=&\frac{1}{\hbar^2}\int_{-\infty}^{\infty}dt {\rm Tr}_R\{\langle E|\tilde{H}_{I}(t)|E'\rangle\langle E'|\tilde{H}_{I}|E\rangle\rho_R^{eq}\}
\end{eqnarray}
and the reflection probability $R(E)$ from the barrier and the transmission probability $T(E)=1-R(E)$ through the barrier.\\
In the transition rates the system-reservoir coupling appears in the interaction picture $\tilde{H}_{I}(t)=e^{i(\tilde{H}_S+\tilde{H}_R)t/\hbar}\tilde{H}_{I}e^{-i(\tilde{H}_S+\tilde{H}_R)t/\hbar}$ with $\rho_R^{eq}={\rm e}^{\beta H_R}/Z_R$ being the equilibrium bath density matrix. Note that the unitary transformation (\ref{eq:unitary}) does not affect the equilibrium density of the bath since $[U_R,H_R]=0$.
The transition rates (\ref{eq:Wq}) can be evaluated explicitly in case of the bilinear system-bath coupling as in (\ref{6:hr}), and one arrives at a golden rule-type of formula
\begin{eqnarray}\label{w}
W_{E,E'}&=&\frac{D_{QQ}(E-E')}{\hbar^2}\left[|Q_{\rm qm}(E',E)|^2+|P_{\rm qm}(E',E)|^2\right]\nonumber\\
&+&\frac{D_{QP}(E-E')}{\hbar^2}2i\Im [Q_{\rm qm}(E',E)^*P_{\rm qm}(E',E)]
\end{eqnarray}
with $Q_{\rm qm}(E',E)\equiv\langle E'|Q|E\rangle$, $P_{\rm qm}(E',E)\equiv\langle E'|P|E\rangle/(\omega_{ d}M)$, and $\Im Z$ denoting the imaginary part of $Z$ . The bath correlation functions $D_{xy}$ correspond to bath forces which couple to $x$ and $y$ according to (\ref{force0}) and (\ref{force}), respectively,
\begin{equation}
\label{bathdensity}
 D_{xy}(E)=\hbar\int_{-\infty}^{\infty}dt \, \tilde{K}_{xy}(t)e^{it E/\hbar}\,.
\end{equation}
In accordance with an effectively Markovian dynamics, we consider a purely ohmic environment with a spectral density $I(\omega)=M\gamma\omega$ in the laboratory frame so that the bath correlation functions in the moving frame are given by \cite{verso_10}
\begin{eqnarray}\label{6:d}
D_{QQ}(E)&=&\tilde{\gamma}M[n_{\beta}(E_F+E) (E_F+E)\nonumber\\
& &+(n_{\beta}(E_F-E)+1)(E_F-E)],\\
D_{QP}(E)&=&i\tilde{\gamma}M[-n_{\beta}(E_F+E) (E_F+E))\nonumber\\
& &+(n_{\beta}(E_F-E)+1)(E_F-E)]\,,\label{6:dp}
\end{eqnarray}
whereas $E_F=\hbar\omega_{ d}$, $\tilde{\gamma}=\gamma/4$ is the effective friction constant in the rotating frame and $n_{\beta}(E)=1/(e^{\beta E}-1)$ is the Bose-Einstein distribution. Interestingly, (\ref{6:d}) displays that physically two channels in the bath are now open and accessible for emission or absorption of quanta, namely, one with energy $E_F+E$ and one with energy $E_F-E$ \cite{verso_10}.

Following the procedure in \cite{verso_09} one arrives with
$
 P(E,t)=\int_{-\infty}^{\infty} P(E',t)\delta(E-E')dE'
$
at an $\hbar$-expansion of (\ref{eq:m}) in the form
\begin{eqnarray}
\label{PE}
\lefteqn{\dot{P}(E,t)=\sum_{k=1}^{\infty}\left(\frac{\partial}{\partial E}\right)^k\frac{1}{k!}\int_{-\infty}^{\infty}d\delta\, W_\delta(E)\ (-\delta)^k}\nonumber\\
&&\hspace{1cm}\times  \frac{R(E)\, P(E,t)}{n(E)}-T(E)\frac{\omega(E)}{2 \pi}P(E,t)
\end{eqnarray}
with $W_\delta(E)=W_{E,E'}$ for $E'-E=\delta$, where $\delta$ is considered as being of order $\hbar$.
The leading order terms in the sum above with $k=1$ and $k=2$  are kept to get the energy diffusion equation for finite transmission in the semiclassical limit, i.e.,
\begin{eqnarray}\label{6:p}
\dot{P}(E,t)&=&\left[\frac{\partial}{\partial E}\left(-\langle\delta \rangle+\frac{\partial}{\partial E} \langle\delta^2\rangle\right)R(E)\right.\nonumber\\
& &\left.-T(E)\frac{\omega(E)}{2 \pi}\right]P(E,t)\, .
\end{eqnarray}
Here, the moments of the energy fluctuations read
\begin{equation}\label{6:delta}
 \langle\delta^k\rangle=\frac{1}{n(E)}\int_{-\infty}^{\infty}d\delta\ \frac{\delta^k}{k!}\, W_{\delta}(E)\,.
\end{equation}

To derive semiclassical transition rates from this diffusion equation one must evaluate the energy momenta up to corrections of order $\hbar$ and also include consistently transmission and reflection coefficients.
Bath induced quantum corrections appear due to bath correlations $D_{xy}$ that enter the transition rate $W_{\delta}(E)$ [see Eq.~(\ref{w})]. Finite tunneling and reflection coefficients appear explicitly in the diffusion equation (\ref{6:p}), but must also be taken into account in the matrix elements $Q_{\rm qm}$ and $P_{\rm qm}$ that determine the system part in the transition rates (\ref{w}).
We note that quantum corrections due to tunneling and reflection are substantial in an energy range $\sim\hbar\omega_{\rm b}$ around the barrier top, where  $R(E_b)\sim T(E_b)\sim 1/2\sim O(\hbar^0)$. They then generate leading corrections in the escape rate that are of order $\hbar$ as well. Quantum corrections that include combinations of finite transmission and bath induced fluctuations are at least of order $\hbar^2$ and can be discarded.
The strategy we follow in the sequel is thus this: in a first step we neglect bath induced corrections and concentrate on the impact of tunneling, while in a second step tunneling is neglected and bath fluctuations are accounted for. Eventually, the corresponding individual energy diffusion equation derived from (\ref{PE}) are combined to capture both phenomena simultaneously.

\section{Semiclassical transition rates}\label{sec5}

In the energy range close to the barrier top, where tunneling dominates in the temperature range considered, the WKB approximation is not applicable because the classical turning points to the left and to the right of the barrier are not sufficiently separated. In the domain around the marginal point, however,  the Hamiltonian $\tilde{H}_S$ can be approximated by an inverted harmonic oscillator  with barrier frequency $\omega_{\rm b}$ [see Appendix (\ref{6:hb})], and one may exploit that the corresponding Schr\"odinger equation can be solved exactly. The proper eigenfunctions are then matched asymptotically (sufficiently away from the barrier top) onto WKB wave functions to determine phases and amplitudes of the latter ones. This way, one obtains
\begin{eqnarray}
\langle E|Q\rangle=\frac{1}{2}\left[\langle E|Q\rangle^-+r(E)\langle E|Q\rangle^+\right]
\end{eqnarray}
with the matrix elements
\begin{equation}\label{eq:Wave}
\langle E|Q\rangle^{\pm}=
N(E)\sqrt{\frac{2\omega(E)}{\pi\p(Q,E)}}e^{\pm\frac{i}{\hbar}S_0(E,Q)\mp\frac{i\pi}{4}}\,,
\end{equation}
containing the action $S_0(E,Q)=\int_{Q_1}^QP(Q',E)dQ'$ of an orbit starting at the turning point $Q_1$ and running in time $t$ towards $Q$ with momentum $ P(Q,E)$ at energy $E$. The complex-valued reflection amplitude $r(E)$ of a parabolic barrier is related to the reflection probability $R(E)=|r(E)|^2$. The normalization follows from $\langle E|E'\rangle=\delta(E-E')$ as
\begin{equation}
N(E)=\frac{1}{\sqrt{\hbar\omega(E)}}\sqrt{\frac{2}{R(E)+1}}\,.
\end{equation}
In case of vanishing transmission, $E<<E_b$ and $R\to 1$, one recovers the standard WKB wave function $N(E)\to \sqrt{1/(\hbar \omega)}$ so that it is possible to use (\ref{eq:Wave}) for all energies, provided the length scale where a parabolic approximation for the barrier applies is much larger than the quantum mechanical length scale $\sqrt{\hbar/M\omega_{\rm b}}$.\\
With (\ref{eq:Wave}) we calculate in the semiclassical limit the transition matrix elements $ Q_{\rm qm}(E',E)$ for finite transmission through the barrier. According to the restricted interference approximation \cite{more_91} we only keep the diagonal contributions of forward/backward waves to obtain
\begin{equation}
Q_{\rm qm}(E',E)=N(E^{\prime})N(E)\left[\left(Q_{\rm scl}^{(\delta)}\right)^*+r(E)r(E^{\prime})^* Q_{\rm scl}^{(\delta)}\right]\, ,
\end{equation}
where
$
Q_{\rm scl}^{(\delta)}(E)\equiv\frac{1}{4}\int_{Q_1}^{Q_2}
 dQ \langle E'|Q\rangle^- Q\langle E|Q\rangle^+\,
$
and $Q_1, Q_2$ denote the left and the right turning points of the periodic orbit with energy $E$, respectively.
The $\hbar$-expansion for the matrix elements is thus found to read \cite{karrlein_97,verso_09}
\begin{eqnarray}\label{eq:expQ}
Q_{\rm scl}^{(\delta)}=Q_{\rm cl}^{(\delta)}+\delta\, \left[\frac{1}{2} \left(Q_{\rm cl}^{(\delta)}\right)^{\prime}+K_Q^{(\delta)}\right]\, ,
\end{eqnarray}
where here and in the sequel the prime $^\prime$ at energy dependent functions denotes the derivative with respect to energy and
\begin{eqnarray}
Q_{\rm cl}^{(\delta)}&=&\frac{\hbar\omega(E)}{2\pi i\delta}\int_{Q_1(E)}^{Q_2(E)}\,dQe^{-i\delta t(Q,E)/\hbar}\\
K_Q^{(\delta)}&=&-\frac{\hbar\omega(E)}{2\pi i\delta}\left.Q(t,E)^{\prime}e^{-i\delta t/\hbar}\right|_{Q_1(E)}^{Q_2(E)}\,.
\end{eqnarray}
Likewise, we calculate the $P$ matrix element
\begin{equation}
P_{\rm qm}(E',E)=N(E^{\prime})N(E)\left[-\left(P_{\rm scl}^{(\delta)}\right)^*+r(E)r(E^{\prime})^* P_{\rm scl}^{(\delta)}\right]\, ,
\end{equation}
with
$
P_{\rm scl}^{(\delta)}(E)\equiv\frac{1}{4\omega_{ d}M}\int_{Q_1}^{Q_2}
 dQ \langle E'|Q\rangle^- \hat{P}\langle E|Q\rangle^+\,
$
and the expansion
\begin{eqnarray}
P_{\rm scl}^{(\delta)}=P_{\rm cl}^{(\delta)}+\frac{\delta}{2} \left(P_{\rm cl}^{(\delta)}\right)^{'}\, ,
\end{eqnarray}
where
\begin{eqnarray}
P_{\rm cl}^{(\delta)}&=&\frac{\hbar\omega(E)}{2\pi\omega_{ d}M i\delta}\int_{P(Q_1,E)}^{P(Q_2,E)}dPe^{-i\delta t(p,E)/\hbar}\,.
\end{eqnarray}
The matrix element in the transition probability (\ref{w}) can then be rewritten as
\begin{eqnarray}\label{eq:w_c}
|Q_{\rm qm}|^2+|P_{\rm qm}|^2&\approx& N^4 \tilde{A} + \delta \tilde{B},\nonumber\\
\Im\, [Q_{\rm qm}(E',E)^*P_{\rm qm}(E',E)]&\approx&N^4\tilde{C} + \delta \tilde{D},
\end{eqnarray}
with coefficients $\tilde{A},\tilde{B},\tilde{C}$, and $\tilde{D}$ specified in Appendix \ref{appA}.

For escape processes near the bifurcation threshold, the energy level spacings of the eigenstates of (\ref{6:hs}) are small compared to $\hbar\omega_{ d}$. Hence, the following approximation of the bath correlations (\ref{6:d}) and (\ref{6:dp}) applies
\begin{eqnarray}\label{6:dqq}
D_{QQ}(E)&=&\tilde{\gamma}M(\kappa-E)+O(\hbar^2)\,,
\end{eqnarray}
where
\begin{eqnarray}
\kappa=\hbar\omega_{ d} \coth(\hbar\beta\omega_{ d}/2)\ ,
\end{eqnarray}
and
\begin{eqnarray}
D_{QP}(E)&\approx&\label{6:dqp}
i\tilde{\gamma}M\hbar\omega_{ d} \frac{E}{|E|}+O(\hbar^2)\,.
\end{eqnarray}
It is important to note that the lowest order in the $\hbar$-expansion of $D_{QQ}$ is of order $\hbar^0$, while that of $D_{QP}$ is of order $\hbar$. Consequently, as we shall see later, the $D_{QP}$-term in (\ref{w}) gives no contribution to the {\em classical} energy diffusion equation [see Eq.(\ref{eq:e_dif_c})].

Now, using (\ref{eq:w_c}), (\ref{6:dqq}), and (\ref{6:dqp}) the $\hbar$-expansion of the transition probability (\ref{w}) takes the form
\begin{equation}\label{6:W2}
W_{\delta}=\frac{1}{\hbar^2}M\tilde{\gamma}N^4\tilde{A}\kappa+\frac{\delta}{\hbar^2}M\tilde{\gamma}\left(-N^4\tilde{A}+\kappa\tilde{B}-2\frac{\hbar\omega_{ d}}{|\delta|}\tilde{C}\right)\,.
\end{equation}
Close to the energy minimum of the stable domains, the energy level spacing $\delta$  is approximately $\hbar\omega_m$, with $\omega_m$ being the minimum local frequency [see App.~\ref{appB}]. Close to the barrier top the energy level spacings vanish so that indeed $\delta\sim\hbar$ and (\ref{6:W2}) is a systematic semiclassical expansion.

\section{Semiclassical escape rates}\label{sec:fin}

Following the discussion at the end of Sec.~\ref{sec4} we start in this Section to consider
 the influence of a finite barrier reflection/transmission in presence of a classical reservoir and then proceed to analyze the impact of bath induced quantum fluctuations for classical reflection/transmission. Both mechanisms are eventually combined in the preceeding Section.

\subsection{Finite transmission}\label{sec:finA}

 With the transition probabilities at hand, the semiclassical diffusion equation follows from (\ref{6:p}) and (\ref{6:delta}) with the semiclassical density of states $n(E)=\frac{1}{\hbar\omega(E)}$ as
\begin{eqnarray}\label{eq:dif}
\dot{P}(E,t) &=&\left[\frac{\partial}{\partial E}\ C(E)\,  \Delta(E)\left(1+\frac{\kappa}{2}\frac{\partial}{\partial E}\right) R(E)-T(E)\right]\frac{\omega(E) }{2 \pi }P(E,t)\,,
\end{eqnarray}
where
\begin{equation}
C(E)=2 \frac{1+R(E)^2}{[1+R(E)]^2}\,
\end{equation}
and
\begin{equation}\label{6:re2}
 \Delta(E)=M\tilde{\gamma}\oint dQ\, \frac{dQ}{dt}+\frac{\tilde{\gamma}}{M\omega_{ d}^2}\oint dP\, \frac{dP}{dt}\,.
\end{equation}
The second term in the \textit{generalized} action (\ref{6:re2}) stems from the $P$ matrix element $|\langle n|P|m\rangle|^2$ in (\ref{w}). We emphasize that no terms originating from the mixed matrix elements $\langle n|P|m\rangle\langle n|Q|m\rangle$ appear in (\ref{w}).

For vanishing transmission ($R=1, T=0$) one recovers from (\ref{eq:dif}) the classical diffusion operator
\begin{eqnarray}\label{eq:e_dif_c}
\dot{P}(E,t)&=&\frac{\partial}{\partial E}\Delta(E) \left(1+\frac{\kappa}{2}\frac{\partial}{\partial E}\right)\frac{\omega(E) }{2 \pi }P(E,t)\, ,
\end{eqnarray}
 which (\ref{eq:e_dif_c}) looks like a classical Kramers equation \cite{kramers_40} with an effective temperature $\kappa$. Further, $\Delta(E)$ corresponds to an energy relaxation coefficient that takes into account the position-position and the momentum-momentum interaction (\ref{introt}) between system and bath.
 As shown in \cite{verso_10} the bath correlation functions $D_{QQ}$ and $D_{QP}$ are associated with two {\em different} effective temperatures due to the fact that a detailed balance condition is not obeyed in the moving frame.
However, since in (\ref{eq:e_dif_c}) [and in (\ref{eq:dif})] the bath correlation function $D_{QP}$ does not play any role, in this regime, it is possible to define the unique effective temperature as $k_{\rm B} T_{\rm eff}=\kappa/2$.

Now, the escape rate is determined by the stationary nonequilibrium distribution $P_{\rm st}(E)$ to (\ref{eq:e_dif_c}), which is associated with a finite flux across the barrier and obeys the boundary conditions $P_{\rm st}=0$ for $E>E_b$ and $P_{\rm st}(E)$ to approach a Boltzmann distribution in the well region. Accordingly, one obtains for high barriers $2 V_b/\kappa\gg 1$ the classical Kramers result
\begin{eqnarray}
\label{classrate}
\Gamma_{\rm cl}&=&\frac{\omega_m \gamma \, \Delta(E_b)}{\kappa\pi}\, {\rm e}^{-2V_b/\kappa}\,
\end{eqnarray}
with the well frequency  $\omega_m$ (\ref{6:wm}).

\begin{figure}[tbp]
\centering
\includegraphics[width=4 in]{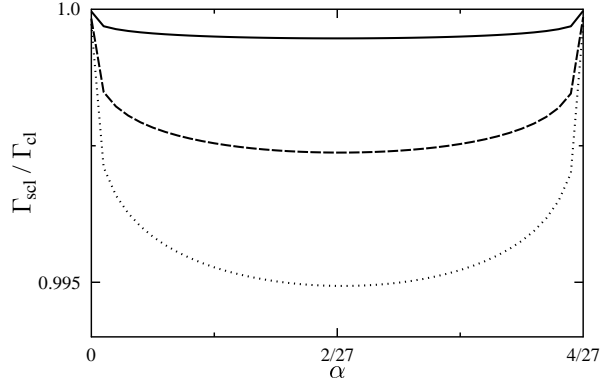}
\caption{Escape rate (\ref{qmrate2}) normalized to the classical rate as a function of the bifurcation parameter $\alpha$ for $\hbar\omega_d\beta/(2\pi)=0.01$ (solid), $\hbar\omega_d\beta/(2\pi)=0.05$ (dashed) and  $\hbar\omega_d\beta/(2\pi)=0.1$ (dotted). For all the lines, we use $\beta L^2\omega_dM\delta\omega=40$, $\delta\omega=0.1\omega_0$ and the dimensionless friction constant $\beta\tilde{\gamma}ML^2\delta\omega=0.1$.} \label{fig:fig6_9}
\end{figure}

The escape rate in the quantum regime including tunneling but no quantum fluctuations in the reservoir can now be evaluated also from (\ref{eq:dif}).
This diffusion equation formally looks like the one already considered in \cite{verso_09} for undriven systems so that we can use the same methods to solve it. In the energy range close the barrier top, where tunneling dominates for the temperature considered here, the approximate Hamiltonian (\ref{6:hb}) leads to the parabolic transmission and reflection probabilities
\begin{eqnarray}
T&=&\frac{1}{1+{\rm exp}\left(-\frac{2\pi (E-V_b)}{\hbar\omega_b}\right)},\\
R&=&\frac{1}{1+{\rm exp}\left(\frac{2\pi (E-V_b)}{\hbar\omega_b}\right)}\,.
\end{eqnarray}
The quantum partition function in the harmonic well region is given by
\begin{equation}
Z_0=\frac{ \kappa}{2\omega_m\hbar}\prod_{n=1}^{\infty}\frac{\nu_n^2}{\nu_n^2+\omega_m^2+\nu_n \gamma}\, ,
\label{partgamma}
\end{equation}
with Matsubara frequencies $\nu_n=\pi n\kappa/\hbar$. For vanishing friction, (\ref{partgamma}) reduces to the known result
$Z_{00}=1/[2{\rm sinh}(\omega_m\hbar/\kappa)]$.
The escape rate follows again from a quasi-stationary nonequilibrium state, this time from the quasi-stationary energy distribution $P_{\rm st}(E)$ of (\ref{eq:dif}), given by
\begin{equation}
\Gamma_{\rm scl}=\int_0^\infty dE\, n(E)\, T(E)\, P_{\rm st}(E)\,.
\end{equation}
This way one gains
\begin{equation}
\label{qmrate}
\Gamma_{\rm scl}=\frac{\sinh(\omega_m\hbar/\kappa)}{(\omega_m\hbar/\kappa)}\, |B|\ \Gamma_{\rm cl},
\end{equation}
with the coefficients
\begin{eqnarray}
B&=&-\frac{1}{4^{\theta}}\frac{_2F_1\left[\frac{1}{2}-\frac{\theta}{2}-a,\frac{1}{2}-
\frac{\theta}{2}+a,1-\theta,-\frac{4}{9}\right]}{_2F_1\left[\frac{1}{2}+\frac{\theta}{2}-a,
\frac{1}{2}+\frac{\theta}{2}+a,1+\theta,-\frac{4}{9}\right]},\\
  a&=&\sqrt{\frac{2\tilde{\gamma} \Delta(E_b)(1-\theta)^2/\kappa+36\theta^2}{8\tilde{\gamma} \Delta(E_b)/\kappa}}\,
\end{eqnarray}
and the abbreviation $\theta=\omega_b\hbar/(\pi\kappa)$.
The first factor in this rate expression captures quantum effects (zero-point fluctuations) in the well distribution, while the second one, $|B|$ describes the impact of finite barrier transmission close to the top.
The latter one can actually prevail and lead to a {\rm reduction} of the escape rate compared to the classical situation due to a finite reflection from the barrier also for energies $E\geq E_b$ (fig.~\ref{fig:fig6_9}).
For a fixed $\tilde{\gamma}$, the expansion of (\ref{qmrate}) for high temperatures is \cite{verso_09}
 \begin{equation}
\label{qmrate2}
\Gamma_{\rm scl}=\Gamma_{\rm cl}\left(1-b_1\theta\right)
\end{equation}
with  $b_1=1.04$ originating merely from the expansion of $B$.
The well partion function leads to corrections  of higher order in $\hbar$.

\subsection{Bath induced fluctuations}\label{sec:finB}

In this section we calculate the impact of the friction terms ${\cal L}_{QP}$ and ${\cal L}_{PQ}$ in (\ref{eq:master_rot})
 on the energy diffusive decay. According to the above strategy, we assume here to have classical transmission and reflection probability, and calculate from (\ref{PE}) the first order $\hbar$ correction to the classical diffusion equation (\ref{eq:e_dif_c}).

For $R=1\,,T=0$ the matrix elements $|Q_{\rm qm}|$ and $|P_{\rm qm}|$ are symmetric with respect to $\delta$, and therefore the terms in the first line of (\ref{w}) do not give contributions of order $\hbar$ to the diffusion equation \cite{linkwitz_91_A,linkwitz_91_B}. The only relevant contributions of order $\hbar$ result from the $\Im [Q_{\rm qm}(E',E)^*P_{\rm qm}(E',E)]$ term.
To calculate it, we must take into account also the next order term in the expansion of the respective bath correlation function, namely
\begin{eqnarray}
D_{QP}(E)&\approx&\label{6:dqp2}
i\tilde{\gamma}M\left[\hbar\omega_{ d} \frac{E}{|E|}+a\hbar|E|\right]+O(\hbar^3)\,
\end{eqnarray}
with $a=\frac{\beta\omega_{ d}-\sinh(\hbar\omega_{ d}\beta)/\hbar}{\cosh(\hbar\omega_{ d}\beta)-1}$. We recall that
energy level spacings are considered as proportional to $\hbar$.
Accordingly, from (\ref{PE}) we obtain the energy diffusion equation,
\begin{eqnarray}\label{diff}
\dot{P}(E,t) &=&\left[\frac{\partial}{\partial E}\, \tilde{\gamma} \left(\Delta -2\hbar\,a\Delta^{(1)}+\left(\Delta \frac{\kappa}{2}+\hbar\,\omega_{ d}\Delta^{(1)}\right)\frac{\partial}{\partial E}\right) \right]\frac{\omega}{2 \pi }P(E,t).
\end{eqnarray}
where
\begin{eqnarray}
 \Delta^{(1)}(E)&=&\frac{8M\pi\tilde{\gamma}}{\hbar}N^4\int_{0}^{\infty}d\delta\,\,\delta^2\Im\left[Q_{\rm cl}^*P_{\rm cl}+Q_{\rm cl}^*P_{\rm cl}^*\right]\,.\label{6:delta1}
\end{eqnarray}
For an explicit evaluation of (\ref{6:delta1}) it is convenient to return to a discrete representation by replacing the energy difference $\delta$ with $\hbar l\omega $ and $\int d\delta $ with $\sum_l\hbar\omega$ so that
\begin{eqnarray}\label{6:delta1b}\Delta^{(1)}(E)&=&\tilde{\gamma}8M\pi\omega\sum_{l=0}^{\infty}l^2\Im\left[Q_{\rm cl}^*P_{\rm cl}+Q_{\rm cl}^*P_{\rm cl}^*\right]\,.
\end{eqnarray}
This expression is correct for low energies, where the spectrum in the wells is discrete, and approximates (\ref{6:delta1}) very accurately for energies near the barrier top.

\begin{figure}[tbp]
\centering
\includegraphics[width=4 in]{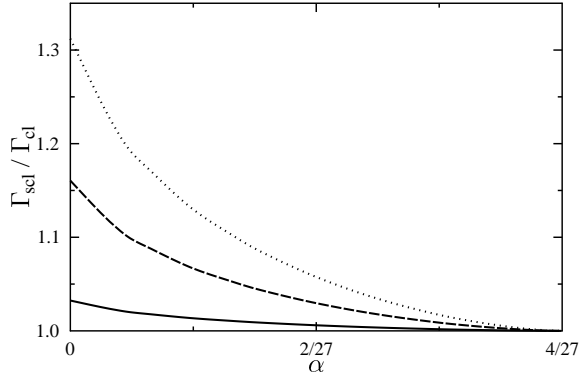}
\caption{Escape rate (\ref{qmrate3}) normalized to the classical rate as a function of $\alpha$ for $\hbar\omega_{ d}\beta/(2\pi)=0.01$ (solid), $\hbar\omega_{ d}\beta/(2\pi)=0.05$ (dashed) and  $\hbar\omega_{ d}\beta/(2\pi)=0.1$ (dotted). For all the lines, we use $\delta\omega=0.1\omega_0$ and $\beta L^2\omega_{ d}M\delta\omega=40$.} \label{fig:fig6_7}
\end{figure}

In order to reveal the effects of the ${\cal L}_{QP}$ and ${\cal L}_{PQ}$ terms, we calculate the rate of escape from (\ref{diff}). Following the standard procedure \cite{kramers_40} one finds
\begin{equation}\label{qmrate3}
\Gamma_{\rm scl}=\Gamma_{\rm cl}\,{\rm e}^{\theta_F b_2}
\,,
\end{equation}
where
\begin{equation}\label{cor}
b_2=\frac{4\pi}{\omega_{ d}\kappa}(a\kappa+\omega_{ d})\int_{E_m}^{E_b}dE\frac{\Delta^{(1)}(E)}{\Delta(E)}\,,
\end{equation}
with $\theta_F=\hbar\omega_{ d}/(\kappa\pi)$. $E_m$ and $E_b$ are the energies of the points (a) and (b), respectively, in fig.~\ref{6:fig_pot}.
The integral in (\ref{cor}) is proportional to the barrier height meaning that (\ref{cor}) is of the order of $V_b\beta$. It thus gives a significant contribution to the escape rate as depicted in fig.~\ref{fig:fig6_7} for various  temperatures.

\section{Discussion}\label{sec6}

In the two previous sections, we have analyzed separately the impact of the two dominant quantum effects on the escape rate including contributions of order $\hbar$. Since corrections due to the combination of the two effects in the transition probabilities are at least of order $\hbar^2$, a full semiclassical diffusion equation up to order $\hbar$ is obtained by simply adding the two results (\ref{diff}) and (\ref{eq:dif}). Hence, we get
\begin{eqnarray}\label{diff2}
 \dot{P}(E,t)=\left[\frac{\partial}{\partial E}\left(\Delta CR-\hbar\,2a\Delta^{(1)}+\Delta C\frac{\kappa}{2}\frac{\partial}{\partial E}R+\hbar\,\omega_d\Delta^{(1)}\frac{\partial}{\partial E}\right) -T\right]\frac{\omega}{2 \pi }P(E,t).
\end{eqnarray}
The leading order quantum corrections to the escape rate are then found as
\begin{equation}\label{qmrate5}
\Gamma_{\rm scl}=\Gamma_{\rm cl}\left[1+\theta_F\left(-b_1\frac{\omega_{\rm b}}{\omega_d}+b_2\right)\right]\,.
\end{equation}
The first correction is negligible when $\omega_{\rm b}\ll\omega_d$, i.e.\  when  $\alpha$ approaches the boundaries of the bifurcation range ($\alpha\to 0$ and $\alpha\to 4/27$).
Interestingly, the two types of quantum fluctuations have opposite effects on the rate expression: while a finite reflection leads for energies above the barrier top to a suppression of the escape probability (fig.~(\ref{fig:fig6_7})), bath induced fluctuations produce an increase (fig.~(\ref{fig:fig6_9})), which typically prevails. The conclusion is thus that in the semiclassical regime finite tunneling through the phase-space barrier does not play an important role, in contrast to quantum fluctuation induced by the reservoir in the moving frame. We recall, that the opposite is true for energy diffusive escape processes over static barriers where tunneling leads to a reduction of the rate \cite{verso_09}.

\begin{center}
\begin{figure}
\centering
\includegraphics[width=4 in]{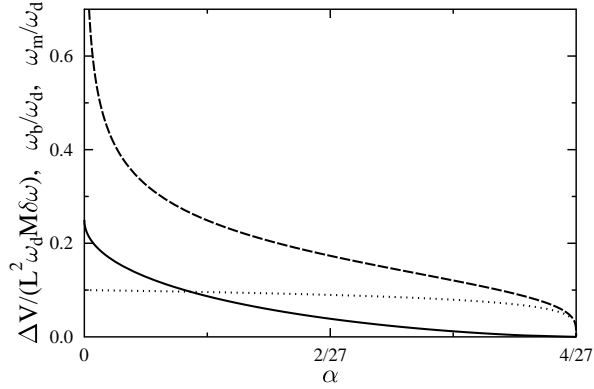}
\caption{The dimensionless barrier height $\Delta V/(L^2\omega_dM\delta\omega)$ (solid line), the well frequency $\omega_m/\omega_d$ (dotted line) and the barrier frequency $\omega_b/\omega_d$ (dashed line) as functions of the bifurcation parameter $\alpha$ for $\delta\omega/\omega_d=0.1$.}\label{6:fig_c}
\end{figure}
\end{center}

It is appropriate to remark  that  the above result is valid  for
\begin{equation}
\Delta V(\alpha)\gg\kappa\gg\hbar\omega_m(\alpha)\,,\hbar\omega_{\rm b}(\alpha)\,
\end{equation}
in order to guarantee the existence of a steady state distribution of a quasi-continuum of thermally smeared states on the one hand and to restrict tunneling to energies close to the barrier top on the other hand. Equivalently,  the range of validity of the rate expression is determined by those values of $\alpha$ which are sufficiently smaller than $\alpha=4/27$ (where $\Delta V\to 0$) [see fig.~(\ref{6:fig_c})].
 \begin{figure}[tbp]
\centering
\includegraphics[width=4 in]{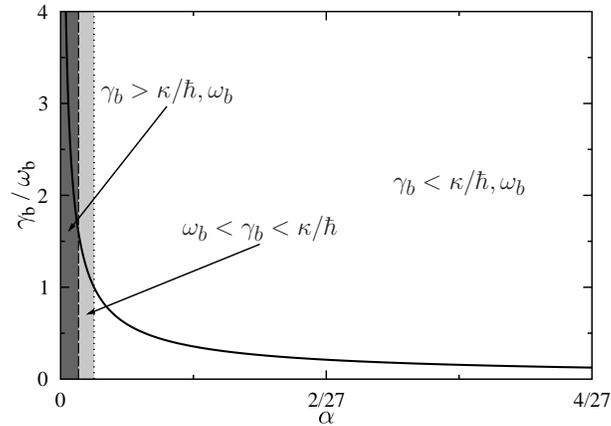}
\caption{Typical behavior of $\gamma_{\rm b}$. For $\alpha$ sufficiently bigger than zero $\gamma_{\rm b}$ is smaller than $\omega_{\rm b}$ and $\kappa/\hbar$. When $\alpha$ goes to zero, $\gamma_{\rm b}$ grows and  achieves a region where it is bigger than $\omega_{\rm b}$ but still smaller than $\kappa/\hbar$. For even smaller $\alpha$, $\gamma_{\rm b}$ is bigger  than $\omega_{\rm b}$ and $\kappa/\hbar$.}\label{fig:fig6_8}
\end{figure}

We have also assumed a weak dissipation compared to the retardation scale $\hbar/\kappa$ of the reservoir and to the time scale of the motion $1/\omega_{\rm b}\,,1/\omega_m$.
However, for $\alpha=0$ the barrier height stays finite, while $\omega_{\rm b}$ goes to zero like $\alpha^{1/4}$ and the effective mass $M_b$ (see Eq.(\ref{6:hb})) tends to infinity like $\alpha^{-1/2}$. The growth of the mass is equivalent to an increase of  the damping which is mostly clearly taken into account when one introduces a rescaled damping constant $\gamma_{\rm b}\equiv\tilde{\gamma}\alpha^{-1/2}$.
Hence, for decreasing $\alpha$, the effective friction $\gamma_{\rm b}/\omega_{\rm b}$ grows as well [see fig.~(\ref{fig:fig6_8})] and, therefore, for sufficiently small $\alpha$ the motion near the barrier becomes overdamped.

In the underdamped regime higher order corrections in the friction appear on the one hand through $\omega_{\rm b}\to \lambda_{\rm b}$ in the factor $B$,
where the Grote-Hynes frequency $\lambda_{\rm b}$ \cite{grote_80} is given by
\begin{equation}
\label{grote}
\lambda_{\rm b}=\sqrt{\frac{\tilde{\gamma}^2}{4}+\omega_{\rm b}^2}-\frac{\tilde{\gamma}}{2}\, .
\end{equation}
On the other hand, the friction dependence of the partition function must be taken into account \cite{verso_09} so that
\begin{equation}\label{qmrate6}
\Gamma_{\rm scl}= \prod_{n=1}^{\infty}\frac{\nu_n^2-\omega_{\rm b}^2}{\nu_n^2-\omega_{\rm b}^2+\nu_n \tilde{\gamma}}\, \Gamma_{\rm cl}\, \left[1+\theta_F\left(-b_1\frac{\lambda_b}{\omega_d}+b_2\right)\right]\,
\end{equation}
with Matsubara frequencies $\nu_n=2\pi n/\hbar\beta$. The motion in the well remains always in the regime $\omega_m/\tilde{\gamma}\ll 1$.

To capture the turnover from weak to strong dissipation, one follows the standard Pollak-Grabert-H\"anggi approach \cite{pollak_86_A,pollak_89_A} and introduces normal-mode coordinates in the parabolic range around the barrier top, where the total system is separable. Then one studies the dynamics of the {\em unstable} normal mode in presence of the coupling to the stable ones due to the potential anharmonicity. For weak dissipation one recovers (\ref{qmrate}) up to leading order.

 With further decreasing $\alpha$ the motion near the barrier becomes overdamped $\gamma_{\rm b}>\omega_{\rm b}$. A classical description applies as long as $\hbar\gamma_{\rm b}<\kappa$ [light gray region in fig.~(\ref{fig:fig6_8})] \cite{ankerhold_01,maier_10} leading to a classical Smoluchowski domain \cite{risken_84}.
Eventually, for very small $\alpha$, friction becomes so strong that  $\hbar\gamma_{\rm b}>\kappa$ [dark gray region in fig.~(\ref{fig:fig6_8})] meaning that the classical Smoluchowski regime turns into the quantum Smoluchowski range \cite{ankerhold_01,maier_10}.
The conclusion of this analysis is that almost everywhere in the bifurcation parameter range $0<\alpha<4/27$ the rate expressions (\ref{qmrate5}) for very weak friction and its extension to somewhat larger fiction (\ref{qmrate6}) are valid.

\section{Conclusions}

In this paper we analyzed the impact of quantum fluctuations on the escape process in case of a dynamical barrier and in presence of a dissipative environment. In the energy diffusive domain of weak friction and higher temperatures a semiclassical procedure allowed to derive effective diffusion equations including leading order quantum effects. It turns out that there are two dominant mechanism for these effects to appear, namely, finite transmission through the barrier and reservoir induced quantum fluctuations in the moving frame. The latter ones dominate by far the deviations to the classical escape rate so that an enhanced  escape probability could experimentally be related to the position-momentum coupling terms that appear in the rotating frame description of the reservoir resting in the laboratory frame.

Interestingly, when the bifurcation parameter tends to zero, the strongly underdamped dynamics turns into an overdamped motion around the barrier top with a friction strength that may even exceed the thermal energy scale. This quantum Smoluchowski domain that so far has only been studied for escape over energy barriers (see e.g.\ \cite{ankerhold_01,maier_10}) will be addressed in a future publication.

\acknowledgements
The authors thank V. Peano for fruitful discussion. Financial support was provided by the German Israeli Foundation (GIF) and the Landesstiftung BW.

\appendix
\section{Coefficients for transition matrix elements}\label{appA}
Here we collect the coefficients appearing in the $\hbar$-expansion of $|Q_{qm}|^2+|P_{qm}|^2$ and $\Im [Q_{qm}^*P_{qm}2]$ in (\ref{eq:w_c}).
One has
\begin{eqnarray}
\tilde{A}&=&\left(\left|Q_{\rm cl}^{(\delta)}\right|^2+\left|P_{\rm cl}^{(\delta)}\right|^2\right)(R^2+1)+R\left(Q_{\rm cl}^{(\delta)^2}+Q_{\rm cl}^{(\delta)^{*^2}}-P_{\rm cl}^{(\delta)^2}-P_{\rm cl}^{(\delta)^{*^2}}\right)\nonumber\\
\tilde{B}&=&\frac{1}{2}\left(\tilde{N}^4\tilde{A}\right)^\prime+\tilde{N}^4\left[-r {r^*}^\prime \left( Q_{\rm cl}^{(\delta)^{*^2}}-P_{\rm cl}^{(\delta)^{*^2}}\right)-r^* r^{\prime}\left(Q_{\rm cl}^{(\delta)^2}-P_{\rm cl}^{(\delta)^2}\right)\right.\nonumber\\
&&\left.+(R^2+1)\left(Q_{\rm cl}^{(\delta)} K^{(\delta)^*}_Q+Q_{\rm cl}^{(\delta)^*} K^{(\delta)}_Q\right)+2 R\left(Q_{\rm cl}^{(\delta)}K_Q^{(\delta)}+Q_{\rm cl}^{(\delta)^*}K^{(\delta)^*}_Q\right)\right]\nonumber\\
\tilde{C}&=&\tilde{N}^4\,\Im\,\left[\left(Q_{\rm cl}^{(\delta)^*}P_{\rm cl}^{(\delta)}-R^2Q_{\rm cl}^{(\delta)}P_{\rm cl}^{(\delta)^*}+R(Q_{\rm cl}^{(\delta)}P_{\rm cl}^{(\delta)}-Q_{\rm cl}^{(\delta)^*}P_{\rm cl}^{(\delta)^*}\right)\right]\\
\tilde{D}&=&\frac{1}{2}\left(\tilde{N}^4\tilde{C}\right)^\prime+\tilde{N}^4\left[r {r^*}^\prime  Q_{\rm cl}^{(\delta)^{*}}P_{\rm cl}^{(\delta)^{*}}-r^* r^{\prime}Q_{\rm cl}^{(\delta)}P_{\rm cl}^{(\delta)}\right.\nonumber\\
&&\left.-R^2P_{\rm cl}^{(\delta)^*}K^{(\delta)}+P_{\rm cl}^{(\delta)}K^{(\delta)^*}+R\left(P_{\rm cl}^{(\delta)}K^{(\delta)}+P_{\rm cl}^{(\delta)^*}K^{(\delta)^*}\right) \right]\,.
\end{eqnarray}

\section{}\label{appB}
Close to its minimum $(Q=Q_m(\alpha),P=0)$, the  Hamiltonian (\ref{6:h}) can be approximated by
\begin{equation}\label{6:hm}
H_{{\rm eff}}^{(m)}=\frac{P^2}{2M_{m}}+V_{{\rm eff}}^{(m)}(Q)\,,
\end{equation}
 where the effective mass is determined by
$$
M_{m}^{-1}=\left.\frac{\partial^2 \tilde{H}_S}{\partial P^2}\right|_m=\frac{\delta\omega}{M\omega_d}\left(1-\frac{Q_m^2}{L^2}\right)
$$
and the effective potential is $V_{{\rm eff}}^{(m)}(Q)=\frac{1}{2}M_m\omega_m^2 Q^2$ with the frequency
\begin{equation}\label{6:wm}
\omega_m=\delta\omega\sqrt{\frac{1-3Q_m^2}{1-Q_m^2}}\,.
\end{equation}
In the same way  it is possible to approximate  the  system Hamiltonian close to the saddle point $(Q=Q_b(\alpha),P=0)$, by
\begin{equation}\label{6:hb}
H_{{\rm eff}}^{(b)}=\frac{P^2}{2M_{b}}+\frac{1}{2}M_b\omega_{\rm b}^2 Q^2
\end{equation}
with $M_{b}^{-1}=\delta\omega/(M\omega_d)\left(1-(Q_b/L)^2\right)$ and
\begin{equation}\label{6:wb}
 \omega_{\rm b}=\delta\omega\sqrt{\frac{1-3(Q_b/L)^2}{1-(Q_b/L)^2}}\,.
\end{equation}

\bibliographystyle{unsrt}
\bibliography{biblio}

\end{document}